%% file: main.tex
\documentclass[aip, pop, amsmath, amssymb, reprint,]{revtex4-1}

\usepackage{graphicx}
\usepackage{dcolumn}
\usepackage{bm}
\usepackage[utf8]{inputenc}
\usepackage[T1]{fontenc}
\usepackage{mathptmx}

\usepackage[dvipsnames]{xcolor}

\newcommand{\eg}{e.g., }
\newcommand{\ie}{i.e., }
\newcommand{\Sec}[1]{Sec.~\ref{#1}}

\newcommand{\Eq}[1]{Eq.~(\ref{#1})}

\newcommand{\Fig}[1]{Fig.~\ref{#1}}

\begin{document}

\title{The Raman gap and collisional absorption}


\author{Ido Barth}
\email{ido.barth@mail.huji.ac.il}
\affiliation{Racah Institute of Physics, The Hebrew University of Jerusalem, Jerusalem, 91904 Israel}
\affiliation{Department of Physics, University of California, Berkeley, California 94720, USA}

\author{Pierre Michel}
\affiliation{Lawrence Livermore National Laboratory, California 94551, USA}


\date{\today}

\input{abstract}

\maketitle

\input{intro}
\input{body}

\bibliographystyle{aipnum4-1}
\bibliography{references.bib}

\end{document}

%% file: abstract.tex
\begin{abstract}

One of the long-standing puzzles observed in many laser-plasma experiments is the gap in the Raman backscattering spectrum. 
This gap is characterized by the absence of backscattered light between some critical wavelength and twice the incident laser wavelength. 
The latter is associated with the absolute Raman instability from the quarter-critical density surface.
Supported by particle-in-cell (PIC) simulations, it is suggested that the gap can result from the collisional damping of the backscattered light. 
A linear analysis of the competition between the Raman growth rate and the damping rate in a non-homogenous plasma predicts the gap's existence and width as a function of the system's parameters. 
The theory is compared with the PIC simulations and past experiments.

\end{abstract}

%% file: intro.tex
\section{Introduction}

The stimulated Raman backscattering (SRS) spectra in laser plasma fusion experiments \eg NIF and Omega, pose a long-standing puzzle.
The wavelength of the backscattered light outside the plasma, $\lambda_1$, is theoretically limited to the range $[\lambda_0, 2\lambda_0]$, where $\lambda_0$ is the incident laser wavelength.
This range results from the Raman resonance conditions
\begin{align} \label{resonance_omega}
    \omega_0&=\omega_1+\omega_2\\
    k_2&=k_0+k_1    \label{resonance_k}
\end{align}
where $\omega_0$, $\omega_1$, and $\omega_2$ are the incident laser, backscattered light, and the electron plasma wave (EPW), respectively, and $k_{0,1,2}$ are the respective wave numbers at the location of the Raman interaction inside the plasma.
$\omega_2$ can be approximated as the plasma frequency, $\omega_p$, by neglecting the thermal correction that changes the SRS spectrum by a few percent for a few KeV electron temperatures. 

However, a typical experimental SRS spectrum is characterized by a gap in the spectrum between some critical wavelength, $\lambda_\text{gap}$ and $2\lambda_0$.\cite{Simon_1986}
This gap, which was not foreseen at the beginning, was first found in experiments, starting with a small amplitude and becoming even more prominent for higher laser intensities.
Since 1985, many theoretical explanations have been suggested, including 
Electron density steepening near the quarter-critical density,\cite{Tanaka_1982,Shepard_1986} 
Large EPW seed driven by a non-Maxwellian distribution with fast electrons,\cite{Simon_1986}
a competition between SRS and stimulated Brillouin scattering (SBS),\cite{Baldis_PRL_1989}
high sensitivity to (damping driven) detuning near the quarter-critical density\cite{Barr_1993},
nonlinear saturation of the plasma wave driven by Langmuir decay instability (LDI),\cite{Drake_1991}
density-dependent diffraction threshold,\cite{Rose_2011}
and back-scatter being overtaken by absolute side-scatter at the relevant densities.\cite{Rosenberg_2020,Michel_2019}

However, the variety and complexity of these explanations hint that the Raman gap effect is not well understood.
Additionally, as far as we know, the Raman gap was never demonstrated within ab initio particle-in-cell (PIC) simulations.

In this paper, we study the effect of collisional damping of the backscattered light on the SRS spectrum as observed outside the plasma.
The idea is that the backscattered light is absorbed in the plasma on its way out.
The amount of absorption depends on the optical depths of the plasma.
Particularly, the larger the backscattered wavelength, the higher density it was scattered from, and the longer the way the backscattered light passes in the plasma and through higher densities.
The latter is because, typically, the incident laser penetrates along the density gradient towards higher plasma densities. 
Therefore, it is anticipated that for some reflecting point, $z_r$, the absorption will be strong enough to cancel the amplification such that the total gain would be zero.
From this point up to the quarter critical density (where the Raman backscattering becomes absolute and thus much higher), no backscattered light is anticipated to exit the plasma, resulting in a gap in the SRS spectrum, as many experiments exhibit.

The paper is organized as follows.
In \Sec{sec2}, we present a linear analysis of the problem and show that the gap can result from collisional absorption.  
A theoretical prediction of the gap location is derived using a single, dimensionless parameter. 
In \Sec{PIC}, we study the Raman gap effect via one-dimension (1D) particle-in-cell (PIC) simulations and discuss the possible explanations in the relevant parameter regimes.
In \Sec{sec4}, we compare our theoretical prediction with past experiments and PIC simulation results.
In \Sec{discussion}, we briefly discuss other theoretical explanations for the Raman gap effect developed in the literature and test their validity against past experimental results and our PIC simulations.
\Sec{conclusions} summarizes the conclusions.

%% file: body.tex
\section{Theory} \label{sec2}
Our analysis is based on the competition between the Raman amplification and the collisional absorption of the backscattered light.
If the former is larger, then backscattered light will be observed, while in the opposite case, all of the backscattered light will be absorbed in the plasma, and no reflected light will be measured.
Since both amplification and absorption depend on the reflection point (defined as the origin of the backscattered light for a given wavelength), one can compare the growth and dumping rates for a wave backscattered from a given location in the plasma.
If the amplification gain is larger than the total damping, SRS backscattered light is expected to be seen in the wavelength associated with the given reflecting point.
On the contrary, if the total damping is larger than the total amplification, no light is expected to be backscattered from the plasma at that wavelength.
The gap effect can be explained by this competition, where the beginning of the gap is determined by the balance between the SRS amplification and the collisional damping, as we will show next for a simple, idealized plasma profile.

\subsection{Density profile}
For simplicity, we consider a slab geometry and a density gradient along the $z-$axis.
The typical ion density, $n_i$ in laser-plasma experiments is generated by isothermal expansion and, therefore, has an exponential profile,\cite{LPI_book} $n_i\sim e^{z/c_s t}$, where $c_s$ is the sound speed and the density gradient is toward the positive $z$ direction.
Since the Raman instability occurs on timescales much shorter than the isothermal plasma expansion, we define a snapshot of the profile
\begin{equation} \label{density_profile}
    n_i = n_\text{min} +\Delta n \; \frac{e^{\alpha z/l}-1}{e^\alpha-1}
\end{equation}
where $l$ is the plasma length, defined as the distance between the two chosen densities, $n_\text{min}$ and $n_\text{max}$. The density difference is $\Delta n=n_\text{max}-n_\text{min}$.
Therefore, 
$l=\alpha c_s t$ for 
\begin{equation}
    \alpha=\ln\left(\frac{n_\text{max}}{n_\text{min}}\right).
\end{equation}
It is noted that in order to get the full SRS spectrum, the value of $n_\text{max}$ must be a little above $n_\text{cr}/4$, and the value of $n_\text{min}$ should be small enough.
Since the simulating of long plasmas is expensive, $n_\text{min}$ in PIC simulations cannot be too small (see \Sec{PIC}), but in the analytical solution below, we will take the limit $n_\text{min}\rightarrow 0$.
A realization of the density profile with the parameters
$n_\text{min}=0.01\, n_\text{cr}$, $n_\text{max}=0.25\, n_\text{cr}$, and $l=1$~mm is illustrated in the inset of \Fig{fig: theory}b.
Surprisingly, as shown below, although the total amplification depends on $l$, the spectral width of the gap does not.

\subsection{Growth rate}
To evaluate the Raman amplification from plasma with a density profile (\ref{density_profile}), we employ the famous Rosenbluth formula for inhomogeneous plasmas providing the convective amplification factor exp$[\Gamma_R]$ for the backscattered light after integration through a resonance region in an inhomogeneous plasma,\cite{Rosenbluth_1972}
\begin{equation} \label{Rosenbluth} 
    \Gamma_R=\frac{\pi\, a_0^2\, k_2^2}{4 k_1}\, L_n
\end{equation}
where the density profile length scale is
\begin{equation}
    L_n=n_e \left(\frac{dn_e}{dz}\right)^{-1}.
\end{equation}
For linear polarization, the dimensionless wave amplitude reads $a_0=\sqrt{2 e^2 \lambda I/( \pi m_e c^5)}\approx 0.02\, \lambda_\mu \sqrt{I_{15}}$, where $\lambda_\mu=\lambda_0/(1\mu m)$ and $I_{15}$ is the incident laser intensity in units of $10^{15} W/cm^2$.
Note that it is justified to use this approximation that depends only on the local gradient because further away from the SRS reflection point, $z_r$, where the resonance condition is exactly fulfilled, the detuning becomes large. 
Therefore, the gain is determined in a small neighborhood of $z_r$.
For Raman backscattering, the wave number amplitude of the plasma wave $k_{2}$ is given by the resonance condition [\Eq{resonance_k}] where
\begin{equation} \label{k_01}
    k_{0,1}=\frac{1}{c}\sqrt{\omega_{0,1}^2-\omega_p^2}    
\end{equation}
are evaluated at the backscattering point, $z_r$, via the (spatially dependent) plasma frequency, 
\begin{equation}
    \omega_p=\sqrt{4\pi e^2 n_e(z_r)/m_e}.    
\end{equation}
Similarly, the backscattered light frequency is determined at the reflecting point, $z_r$, via the approximated \Eq{resonance_omega},
\begin{equation}\label{w_1}
    \omega_1=\omega_0-\omega_p(z_r)
\end{equation}
while the incident laser frequency, $\omega_0$, does not depend on the plasma parameters.

\begin{figure}[tb]
    \includegraphics[clip, trim=2.cm 1.2cm 2.0cm 2cm,
    width=1.0\linewidth]{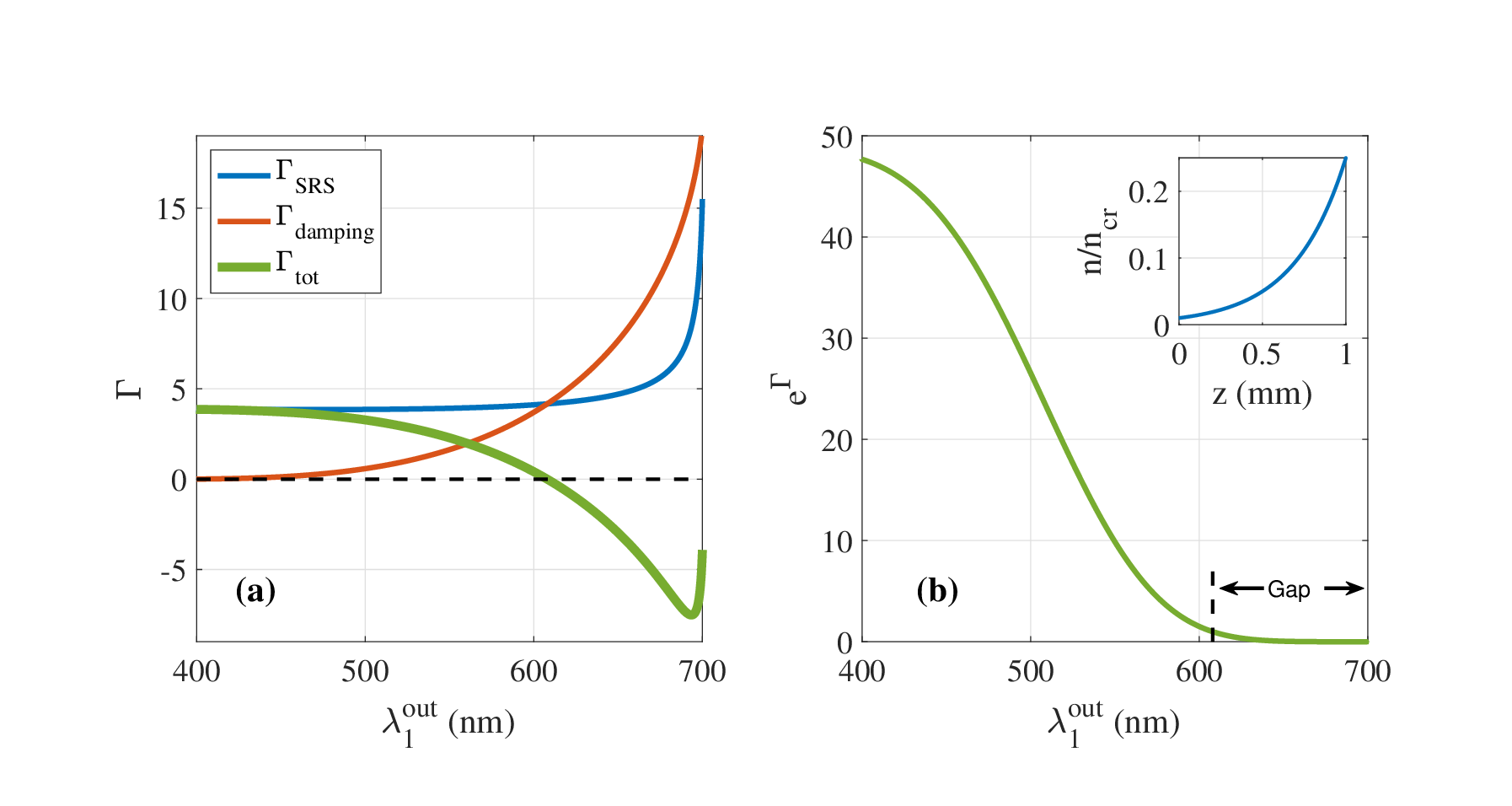}
    \caption{(a): SRS gain from the Rosenbluth formula in \Eq{Rosenbluth} (blue), collisional damping loss from \Eq{damping} (red), and the difference between the two, $\Gamma_\text{tot}$, (green) that crosses zero (dashed black) at the gap location. (b): The total gain as a function of the backscattered light wavelength outside the plasma. The inset is the electron density profile of \Eq{density_profile}.}
    \label{fig: theory}
\end{figure}

\subsection{Collisional absorption}

The amount of absorption that light experiences	on its way out of the plasma, $\exp(-\Gamma_d)$, can be estimated by integrating the local damping rate over the distance between the reflecting point, $z_r$, and the plasma edge, $z=0$, (\ie where $n_i=n_\text{min}$)
\begin{equation} \label{damping}
    \Gamma_d=\int_0^{z_r} \kappa\, dz.
\end{equation}%
The simplest model for the local damping rate of light in plasma is given by\cite{Kruer_book}
\begin{equation}
     \kappa = \frac{\omega_1}{c} \Im [\epsilon] \approx \frac{\omega_p^2 \nu_{ei}}{\omega_1^2 v_g}.
\end{equation}%
Here, $\Im[\epsilon]$ is the imaginary part of the electric permittivity, $v_g=c^2 k_1/\omega_1$ is the group velocity of the backscattered light, and 
\begin{equation}
	\nu_{ei}= \frac{Z\, \ln \Lambda }{3 (2\pi)^{1.5}}\; \frac{\omega_p }{n_e \, \lambda_D^3 }
\end{equation}%
is the electron-ion collision rate\cite{Kruer_book} , which depends on the temperature through the Debye length, $\lambda_D=\sqrt{k_B T /4\pi e^2 n_e}$. 
The Coulomb logarithm can be estimated by\cite{LPI_book} 
\begin{equation}
    \ln{\Lambda} = \ln{\sqrt{\frac{b_\perp^2+b_{max}^2}{b_\perp^2+b_{min}^2}}},
\end{equation}%
where, $b_{\perp}=\frac{Z^{*} e^2}{4\pi \epsilon_0 m_e v_{e}^2} $, $b_{max}=\frac{v_e}{\omega}$, and $b_{min}=\frac{\hbar}{2 m_e v_e}$.
Here, $v_e$ is the thermal velocity of the electrons (estimated in the low-field limit\cite{LPI_book}), $Z^{*}=\left<Z^2\right>/\left<Z\right>$ is the effective ionization number, $e$ is the electron charge, and $m_e$ is the electron mass.
These quantities can be estimated from the diagnostics of a given experiment.
Therefore, if the experimental parameters are measured or estimated with sufficient accuracy, one can predict the location of the beginning of the gap in the SRS spectrum by equating
\begin{equation}\label{Gamma}
    \Gamma_{\rm R}=\Gamma_d,
\end{equation}
and solving (numerically) for the $\omega_1$.
Because the diagnostics are placed outside the plasma, the location of the beginning of the Raman gap in the wavelength $x-$axis is $\lambda_1^{\rm out}=2\pi c/\omega_1$.
For the example presented in \Fig{fig: theory}, the gap location is found to be at $608$~nm, which, in the left panel, it is associated with the crossing point of $\Gamma_{\rm tot}=\Gamma_{\rm R}-\Gamma_d$ and the $x$-axis.
It is also indicated in the right panel of the figure by a vertical dashed line.

\subsection{Single dimensionless parameter}
It is notable that \Eq{Gamma}, which defines the location of the gap, can be simplified by isolating all the physical parameters except the laser frequency, $\omega_0$, into one side of the equation, leaving on the other side only an integral expression that, for a given $\omega_0$, depends only on the reflecting point $z_r$,
\begin{equation}\label{all_togather}
    \frac{I \left(k_B T\right)^{1.5} }{Z^{*}\ln\Lambda} = 
    \frac{2k_1\, m_e^{2.5}\, c^4}{3\left(2\pi\right)^{1.5} k_2^2\, \lambda_0^2 \, e^2}
    \int_{0}^{z_r} \frac{dz}{\omega_1^2v_g n_e}.
\end{equation}
It is noted that for a given $\omega_0$, the electron density at $z=z_r$ determines all other frequencies and wave numbers via the SRS resonance conditions [Eqs.~(\ref{resonance_omega}-\ref{resonance_k})] and Eqs.~(\ref{k_01},\ref{w_1}), 
\begin{align}
    \omega_p &= \omega_0\sqrt{n} \\
    \omega_1 &= \omega_0\left(1-\sqrt{n}\right) \label{w1_of_n} \\
    k_0 &= \frac{\omega_0}{c}\sqrt{1-n}\\
    k_1 &= \frac{\omega_0}{c}\sqrt{\left(1-\sqrt{n}\right)^2-n}\\
    k_2 &= \frac{\omega_0}{c}\left(\sqrt{1-n} + \sqrt{\left(1-\sqrt{n}\right)^2-n}\right)
\end{align}
where, $n=n_e(z_r)/n_\text{cr}$ is the dimensionless electron density at $z=z_r$.
Therefore, by solving \Eq{all_togather} for $z_r$, one can find the spectral gap location, $\lambda_\text{gap}=2\pi c /\omega_1$, which is the wavelength outside the plasma of the light that was backscattered from $z=z_r$.

By employing the definitions of all the parameters in the right-hand-side (RHS) of \Eq{all_togather}, the laser frequency can also be isolated such that the RHS will depend only on $z_r$. 
At the same time, the left-hand-side (LHS) includes an additional term of $\lambda_0^3$.
Therefore, it is constructive to define one dimensionless parameter that includes all of the physical parameters in the problem,
\begin{equation} \label{xi_definition}
    \xi=\frac{I_{15} T_{\rm eV}^{1.5} \lambda_\mu^3} {Z \ln\Lambda},
\end{equation}
where $I_{15}=I/10^{15} W cm^{-2}$, $T_{\rm eV} = k_B T/eV$, and $\lambda_\mu=\lambda_0/\mu m$.
Now, \Eq{all_togather} can be rewritten in a dimensionless form,
\begin{equation} \label{xi_final}
    \xi=2996 \frac{\left(1-\sqrt{n}\right)^2 \sqrt{(1-\sqrt{n})^2-n}}{\left(\sqrt{1-n} + \sqrt{ (1-\sqrt{n})^2 -n}\right)^2} \; J,
\end{equation}
where,
\begin{equation}
    J=\int_{y_0}^{y_1}\frac{y}{\sqrt{1-y}}\, dy
\end{equation}
with $y=\frac{\omega_0^2}{\omega_1^2}n$.
The integration limits are $y_0=\frac{\omega_0^2}{4\omega_1^2}e^{-\alpha/2}$ and $y_1=(1-\frac{\omega_0}{\omega_1})^2$.
Note that $y_0\rightarrow 0$ when $n_\text{min}\rightarrow 0$ because then $\alpha\rightarrow\infty$.
Assuming this limit, the analytical solution of the dimensionless integral reads
\begin{equation}
    J=\frac{1}{6}\cos(3\theta_1) -\frac{3}{2}\cos(\theta_1) + \frac{4}{3}
\end{equation}
where
\begin{equation}
    \theta_1 = \sin^{-1}\left(1-\frac{1}{1-\sqrt{n}}\right)
\end{equation}
Notably, we obtained in \Eq{xi_final} a relation between the density at the gap location, $n=n_\text{gap}$, and the dimensionless parameter, $\xi$, which is determined by the system's parameters in \Eq{xi_definition},
\begin{equation} \label{n_gap}
    \xi=f(n_\text{gap}).
\end{equation}
This relation is depicted in a solid blue line in \Fig{fig: experiments} (left $y-$axis).
By numerical inverting \Eq{n_gap}, one can find the theoretical gap location in terms of the plasma density as a function of the systems' dimensionless parameter $n_\text{gap}=n_\text{gap}(\xi)$.
However, this is a multi-valued function, so we omit the higher branch (near the quarter critical density, $n_{cr}/4$) because the group velocity of the backscattered light becomes very small, and the SRS interaction is nearly absolute.
Finally, for a given laser frequency, $\omega_0$, the wavelength at the gap, $\lambda_\text{gap}$ can also be calculated from \Eq{n_gap} and \Eq{w1_of_n}.
The right $y-$axis of \Fig{fig: experiments} is associated with the theoretical prediction for the spectral gap location in terms of $\lambda_\text{gap}/\lambda_0$.   


\section{PIC simulations} \label{PIC}

\begin{figure}[tb]
    \includegraphics[clip, trim=1cm 1.2cm 1.5cm 1.cm,
    width=1.0\linewidth]{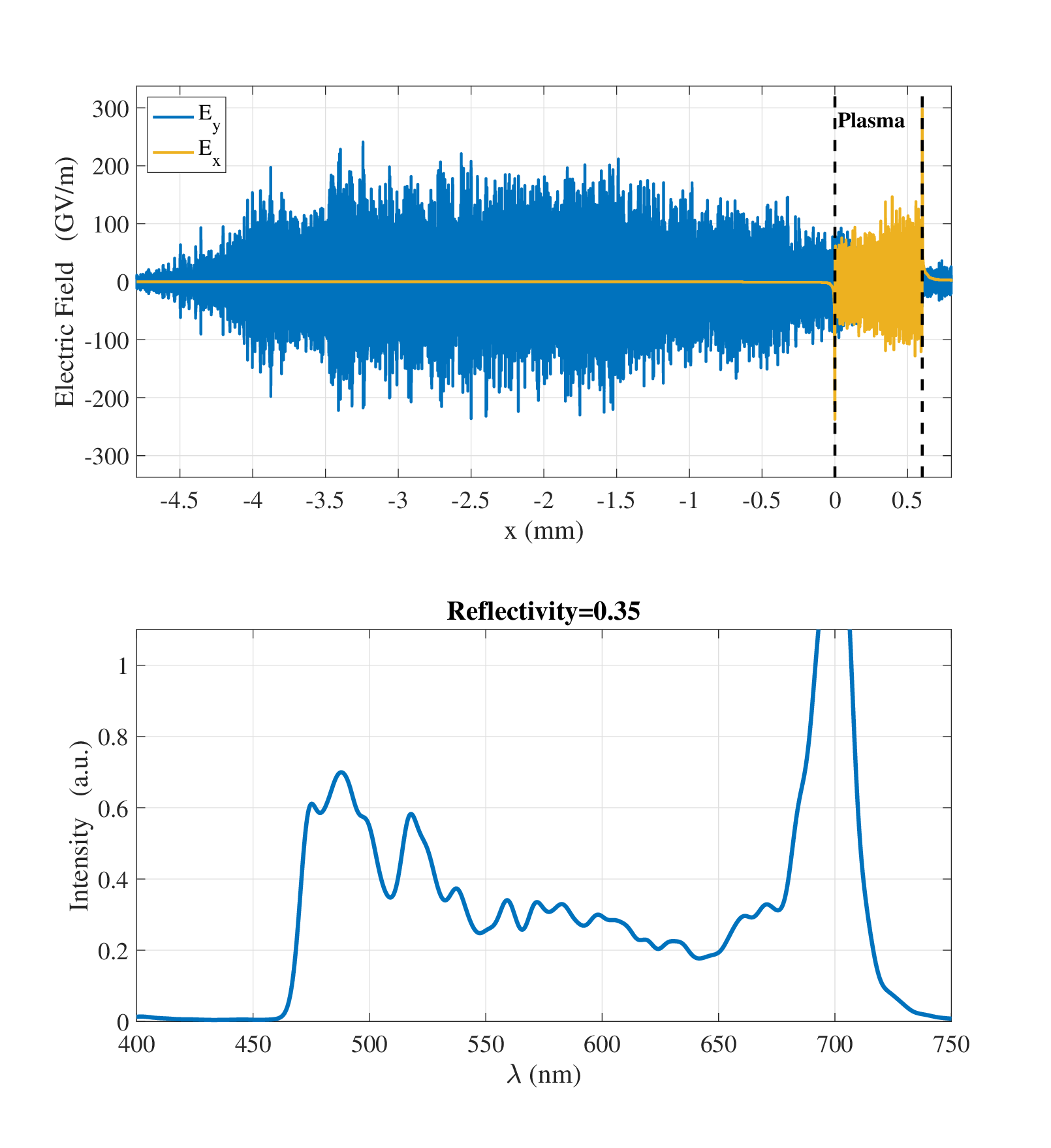}
    \caption{PIC simulation results for negligible collisions ($\ln \Lambda =1$). Upper panel: A snapshot of the transverse electric field, $E_y$ (blue), and the longitudinal electric field, $E_x$ (yellow), at a time when the whole incident laser pulse has passed the plasma (denoted in dashed black lines) from left to right. Lower panel: The spectrum of the backscattered light between $-4.8$~mm and zero of the upper panel. No noteworthy gap is observed in the Raman spectrum in this case.}
    \label{fig: PIC_no_gap}
\end{figure}

\begin{figure}[tb] 
    \includegraphics[clip, trim=1cm 1.2cm 1.5cm 1.0cm,
    width=1.0\linewidth]{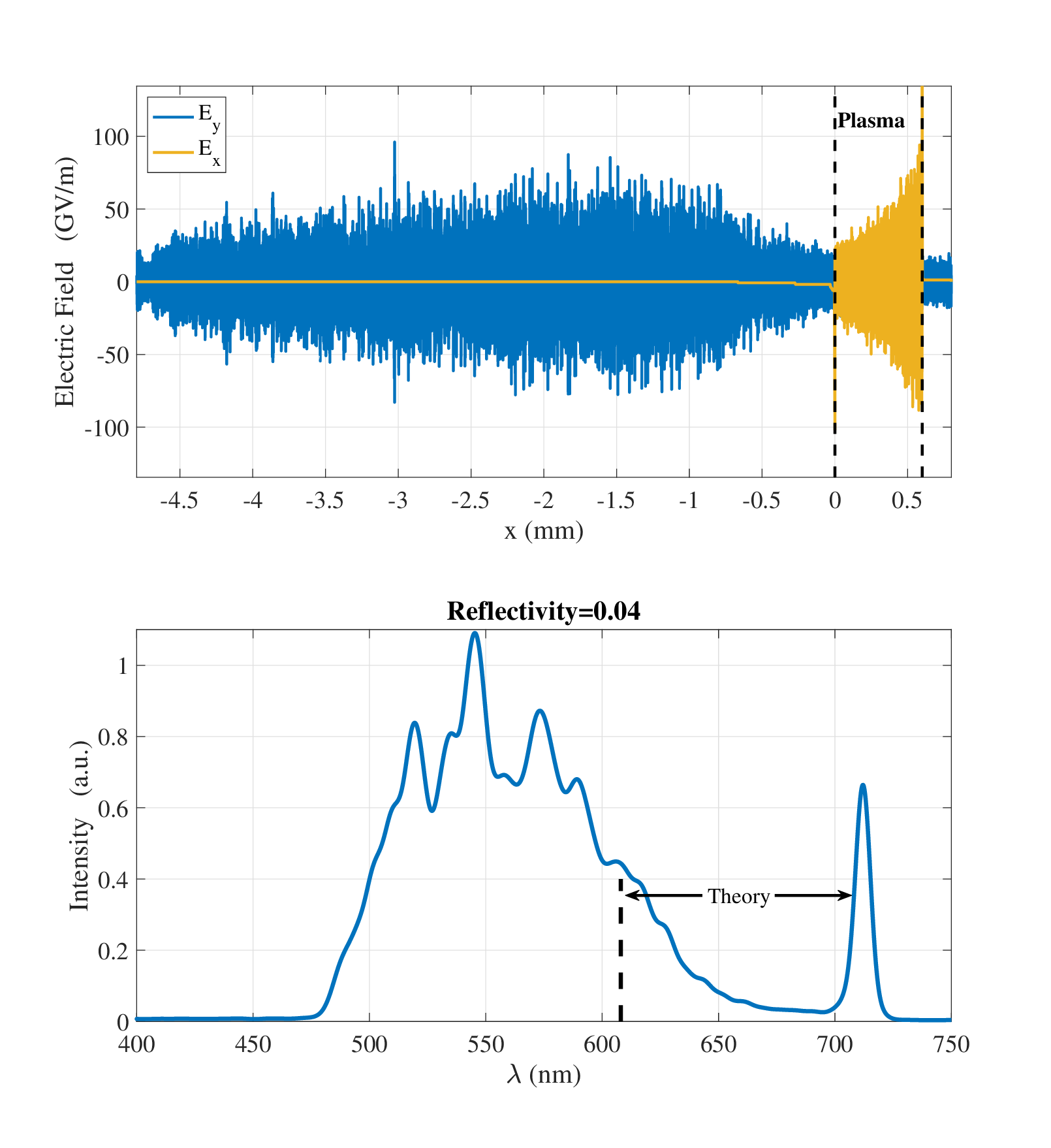}
    \caption{PIC simulation results for high collisions ($\ln \Lambda =80$). Panels are the same as in \Fig{fig: PIC_no_gap}. A gap in the SRS spectrum is observed between about $640$~nm and the (thermally shifted) absolute SRS at about $710$~nm. The theoretical location of the lower end of the gap, $608$~nm, is denoted by a vertical dashed line.}
    \label{fig: PIC_gap}
\end{figure}

To study the Raman gap effect, we have run 1D PIC simulations using the code EPOCH \cite{epoch} with $250$ cells per $\mu$m and $50$ particles per cell.
It is noted that in addition to the physical noise, PIC simulations also exhibit numerical noise, which depends on the resolution.
Therefore, the total reflectivity is not well reproduced, but relative effects can be investigated better within a fixed resolution.
In the simulations, we consider typical physical parameters of inertial confinement fusion (ICF) experiments as follows:
The laser wavelength and intensity were $\lambda_0=0.351 \, \mu$m and $I_0=2.5\times10^{15}$~ W/cm$^2$;
the pulse duration was $\tau=12$~ps;
both electron and ions ($Z=1$) density profiles ($n_e=n_i$) were exponential as defined in \Eq{density_profile} with  $n_{\rm min}=0.05\, n_{\rm cr}$, $n_{\rm max}=0.27\, n_{\rm cr}$, and $l=0.6$~mm;
the electron temperature was $T_e=1000$~eV while the ions were immobile.

In \Fig{fig: PIC_no_gap} (upper panel), we plot a snapshot of the electric field, $E_y$ of the backscattered light (blue) and the longitudinal electric field, $E_x$, of the EPW inside the plasma, for negligible collisions ($\ln\Lambda=1$). 
Dashed lines depict the plasma boundaries.
The incident laser pulse propagates from left to right, and the reflected light propagates from right to left.
Thus, at the snapshot time, the left part of the figure represents the reflected light while the small right part (right to the plasma) is the transmitted pulse.
The spectral analysis of the reflected part is shown in the lower panel of \Fig{fig: PIC_no_gap}.
Notably, it does not exhibit the gap effect as the SRS intensity around $600-700$~nm is similar to the intensity at smaller wavelength, \ie within the range of $50$ percent from the averaged intensity at $500-600$~nm.
 
On the contrary, in \Fig{fig: PIC_gap}, we present the results of an identical PIC simulation but with $\ln \Lambda=80$, \ie stronger collisions.
First, we note that the total SRS reflectivity in the case with collision, $4$ percent, is significantly smaller than that without collisions, $35$ percent, and much more similar to the experimental values of SRS reflectivity.
Second and most notably, the spectrum exhibits a prominent gap between $630$~nm and the absolute SRS at $710$~nm (that agrees with the thermal shift), in which the SRS intensity is about $95$ percent less than the average intensity at the range of $510-590$~nm.
Finally, the theoretical prediction for the gap location for this simulation's parameters is $608$~nm (denoted by a dashed black line), which reasonably agrees with the gap location in the simulation.

The comparison between these two simulations suggests 
that collisional absorption is a plausible explanation for the Raman gap effect.
Moreover, other explanations for the effect can be tested by their ability to explain the results of the 1D PIC simulations presented here.
We will remember this argument when discussing the alternative explanations in \Sec{discussion}. 

\begin{figure}[tb] 
    \includegraphics[clip, trim=1.5cm 0.1cm 1.8cm .5cm,
    width=1.0\linewidth]{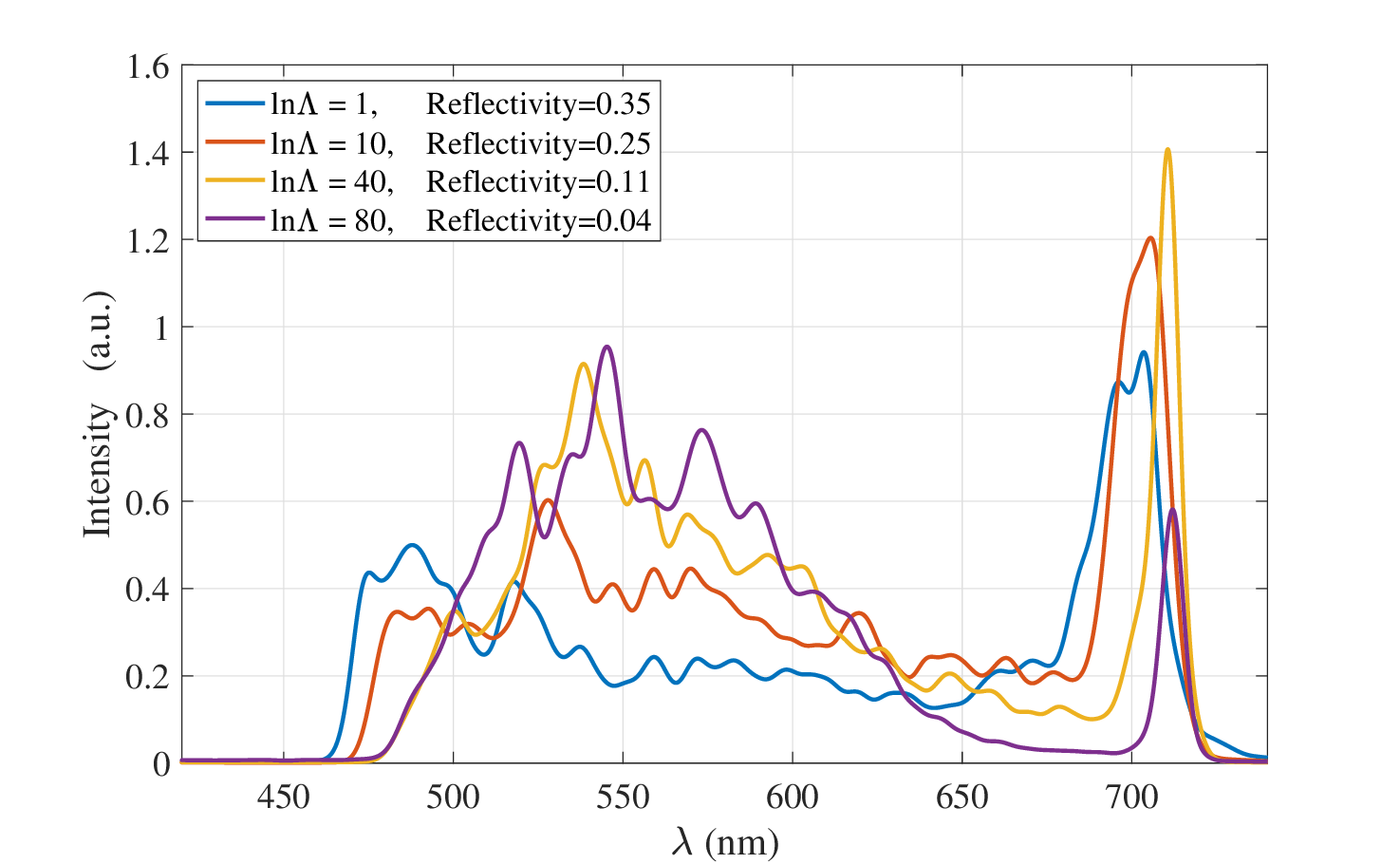}
    \caption{Illustration of the evolution of the Raman gap with collisionality. The SRS spectra of four PIC simulations with different values of $\ln\Lambda, \{1,10,40,80\}$, are presented. For better visualization, each spectrum is normalized by its total SRS reflectivity (see legend).}
    \label{fig: lnLambda_dependency}
\end{figure}

To further illustrate how the gap evolves when increasing the collisionality, we repeat the same simulation with two intermediate values of $\ln\Lambda$ and plot the spectra in \Fig{fig: lnLambda_dependency}.
Each spectrum in the figure is normalized by its associated total SRS reflectivity, as denoted in the legend.
It can be seen that as the collisionality (\ie the value of $\ln\Lambda$) increases, the ratio between the SRS intensity in the range of $[640-700]$~nm and the intensity in the range of $[500-600]$~nm decreases.

We conclude that the introduction of collisions in the PIC simulations yields the appearance of the Raman gap, in agreement with the linear analysis derived in \Sec{sec2}.
In the next section, we will test the theoretical prediction for the location of the gap against a few past experiments.

\section{Comparison with past experiments} \label{sec4}

To validate our theory against past experiments, we have analyzed a few published experimental Raman spectra.
To this end, a reasonable estimation of the physical parameters in each experiment is required.
The relevant parameters for the analysis are the electron temperature, the laser intensity, and the degree of ionization. 
Also, we implicitly assumed an exponential density profile.
However, in cases where the density profile is different but known, the theoretical gap location can be easily calculated by numerically integrating \Eq{all_togather} over the measured density profile.

Unfortunately, although the Raman gap effect was observed in many LPI experiments, the diagnostic of the laser or the plasma parameters was poor in many of them.
Particularly in most of the relevant publications, the temperature of the electron temperature was not measured or, at least, not reported in the paper.
Besides, even when measurements or estimations for values of the physical parameters are given, the uncertainty in the parameters' values is usually large.
These uncertainties must be reflected by error bars in the graph.
In the absence of reported estimation for the error bars, we arbitrarily assumed an error of $10-20$ percent, where we note that the uncertainty in the degree of ionization, $Z$, is relevant only in experiments with a gold target.

The location of the gap was extracted from the figures in the papers by a rough estimation of the location where the SRS spectrum decreased by about $90$ percent from the averaged value at wavelengths below the gap.
We employ this definition because it is consistent with the definition of the theoretical location of the gap, $\Gamma_{tot}=0$, meaning no SRS amplification of the noise.

Because of the quality of the figures in the literature, the estimation was done by eye and a ruler.
For each spectrum, we associate some error estimated from the spectral curve's slope.
The data extracted from the literature, the error bar associated with the physical parameters, and our extracted gap locations are summarized in Table \ref{tab:table2}.

\begin{figure}[b]
    \includegraphics[clip, trim=.9cm 2.3cm 0.0cm 3.8cm,
    width=1.0\linewidth]{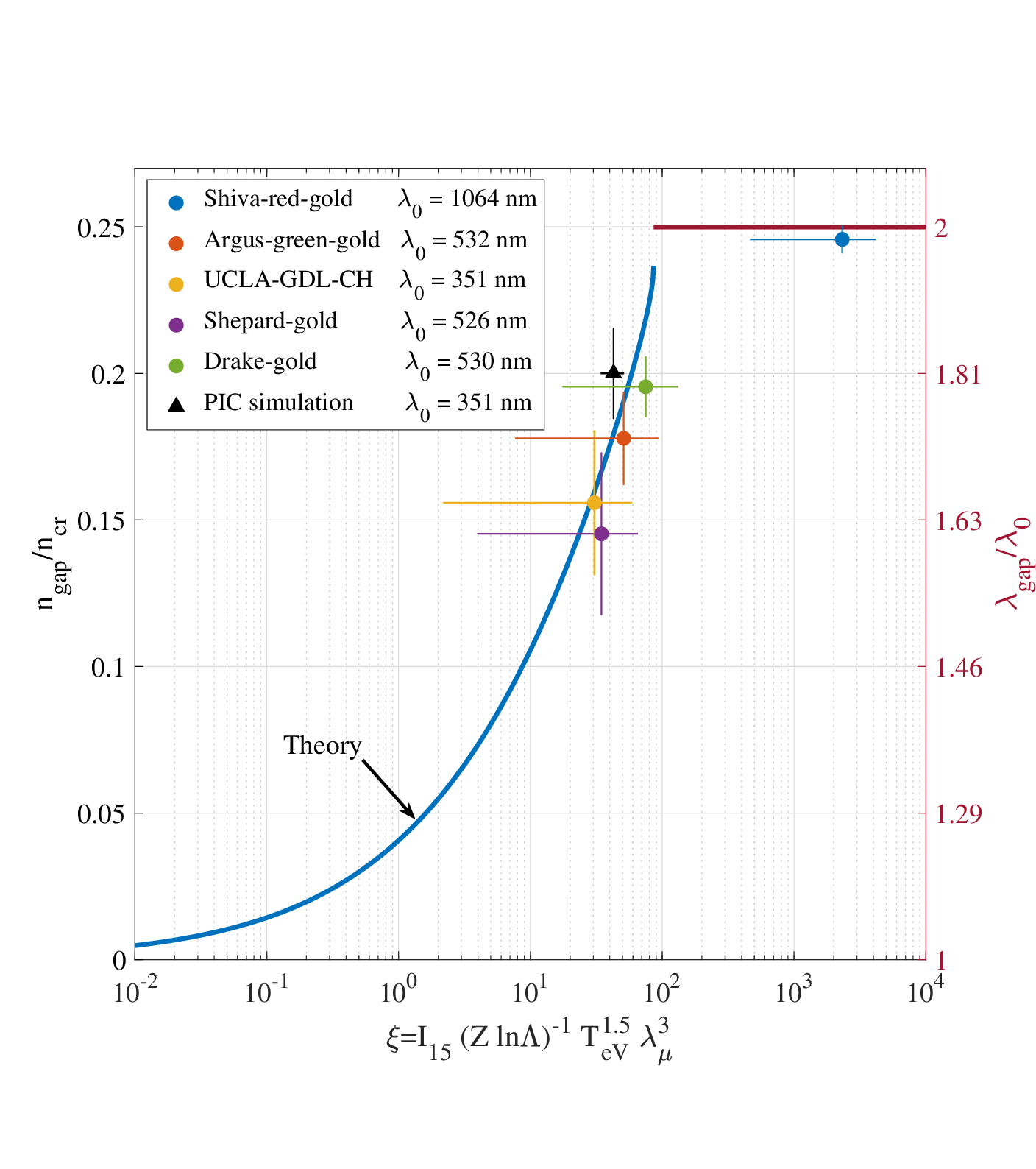}
    \caption{The theoretical gap location as a function of the dimensionless parameter, $\xi$ (solid blue line), in comparison with several past experiments (colored circles) and our PIC simulation example (black triangle). Also denoted the gap's upper limit at twice the laser wavelength \ie for quarter critical density (solid red line). The left $y-$scale is the gap density normalized by the critical density, while the right $y-$scale is the associated normalized wavelength of the backscattered light.}
    \label{fig: experiments}
\end{figure}

The results are presented in \Fig{fig: experiments}.
Each experiment is denoted by a different color listed in the legend.
The theoretical location of the gap [\Eq{n_gap}] is denoted by a solid blue line.
The range of values of the dimensionless parameter, $\xi$, where the theory predicts there should not be a gap, is denoted by a solid horizontal red line associated with the quarter critical density and a gap "location" of $2\lambda_0$.
Interestingly, there is one experiment ("Shiva-red-gold") with parameters in this regime, and indeed, only a tiny gap was observed in the SRS spectrum in this experiment.
A good agreement between the experimental dots and the theoretical curve can be seen.
In addition, the PIC simulation result is denoted by a black triangle and agrees well with the theory.

\begin{table*}
\caption{\label{tab:table2} List of the parameters in past experiments and PIC simulations. }

\begin{ruledtabular}
\begin{tabular}{ccccccccc}
 Name & Source & Target & $\lambda_0$~[nm] & $I_{15}$ &$Z^{*}$& $T_{\rm KeV}$ &$\lambda_\text{gap}$~[nm] &  \\
\hline

Shiva-red & Ref.~\onlinecite{Simon_1986} (Fig~6)
    & Gold\footnotemark[1] & $1064$ & $2\pm 1$ &  $50\pm5$& $5\pm1$ & $2110\pm 20$&  \\

Argus-green & Ref.~\onlinecite{Simon_1986} (Fig 8)
    & Gold\footnotemark[1] & $532$ & $1\pm 0.5$ &  $50\pm5$& $2\pm0.5$ & $920\pm 30$&  \\

UCLA-GDL & Ref.~\onlinecite{Simon_1986} (Fig~14)
    & CH\footnotemark[2] & $351$ & $1\pm 0.5$ &  $4.75\pm0$ & $0.7\pm0.3$ & $580\pm 30$&  \\

Shepard & Ref.~\onlinecite{Shepard_1986} 
    & Gold\footnotemark[1] & $526$ & $0.7\pm 0.3$ &  $50\pm5$ & $2\pm1$ & $850\pm 50$&  \\
    
Drake & Ref.~\onlinecite{Drake_1991}
    & Gold\footnotemark[1] & $530$ & $0.9\pm 0.3$ &  $50\pm5$ & $3\pm1$ & $950\pm 20$&  \\

PIC simulation & This work (\Fig{fig: PIC_gap})
    & Immobile ions\footnotemark[3] & $351$ & $2.5$ &  $1$ & $1$ & $635\pm 20$&  \\
    
\end{tabular}
\end{ruledtabular}
\footnotetext[1]{For golds target, we assumed partial ionization of $Z=50$ with $10\%$ error.}
\footnotetext[2]{For CH targets, we assumed fully ionized plasma without uncertainty.}
\footnotetext[3] {Numerical collisions strength was set to  $\ln{\Lambda}=80$. The numerical error is estimated as $20\%$. Other numerical parameters can be found in \Sec{PIC}.}
\end{table*}


\section{discussion} \label{discussion}
The Raman gap that was observed in many experiments poses a long-standing puzzle.
In the literature, we found seven different theories to explain the effect.
While we cannot discern the level of verisimilitude claimed in each explanation, we point out that it is reasonable to check if they can explain the gap effect observed in our PIC simulation results.
First, as mentioned in \Sec{PIC}, the simulations were one-dimensional and considered immobile ions but exhibited the Raman gap effect.
Therefore, theories that are based on the dynamics in more than one dimension\cite{Rose_2011,Rosenberg_2020} or involve interaction with ion-acoustic waves\cite{Baldis_PRL_1989,Drake_1991} (\ie ion motion) can not explain the gap effect in the simulations.
Second, PIC simulations are costly, and thus, the laser pulse duration in the simulation was only $12$~ps.
Although this timescale is enough for studying the Raman backscattering, it is not sufficient for developing a population of fast electrons (especially without the TPD process, which is a 2D effect).
Therefore, the simulations do not support the theory based on enhanced EPW seed driven by fast (non-Maxwellian) electron distribution\cite{Simon_1986}.
Third, the theory based on high sensitivity to detuning near the quarter critical density\cite{Barr_1993} considered extreme density gradients, \ie very short plasmas (about $100$ wavelength only).
The plasma length in the PIC simulations was too long ($600 \mu$m) to be adequately comprehended by this approach.
Finally, the idea of electron density profiles that steepen near the quarter critical density such that Raman backscattering near $2\lambda_0$ is substantially reduced\cite{Tanaka_1982,Shepard_1986} may require a tailored density profile to explain the observed Raman gap effect. 
In contrast, for typical density profiles in our PIC simulations, the gap effect was not observed in the absence of collisions but was observed when collisions were introduced.

It is important to emphasize that we do not claim that the aforementioned theories do not contribute to forming the gap in the SRS spectrum.
We argue that, on the one hand, the effect in both experiments and PIC simulations can be understood as a result of the competition between SRS amplification and collisional absorption.
On the other hand, other theories, most of which are much more complicated, might explain the experiments but not the PIC simulations.
Additionally, our theoretical prediction for the gap width is in reasonable agreement with past experiments without the need for other supportive effects.

\section{conclusions} \label{conclusions}

In conclusion, we developed a theory for the Raman gap effect based on the competition between the SRS growth rate in inhomogeneous plasmas and the collisional absorption of the backscattered light on its way out of the plasma.
The gap formation was observed in a series of PIC simulations where only the collisionality changes. 
The location of the gap is calculated via linear analysis and exhibits a reasonable agreement with several past experiments and our PIC simulations.
Previous theoretical explanations suggested in the literature for the Raman gap can not explain the PIC simulation results.
Therefore, we conclude that the Raman gap can often result from the collisional absorption of the backscattered light without the support of other explanations.

\section*{Data Availability Statement}
The data that support the findings of this study are available from the corresponding author upon reasonable request.

\section*{Acknowledgments}
This work was performed under the auspices of the U.S. Department of Energy by Lawrence Livermore National Laboratory under Contract DE-AC52-07NA27344, and was supported by the PAZI Foundation, Grant No. 2020-191.
The computations in this paper were run on the ICPL cluster at the Hebrew University of Jerusalem.
One of the authors (IB) would like to thank Jonathan Wurtele for his hospitality at UC Berkeley, where this work was mainly conducted.

%% file: main.bbl
\begin{thebibliography}{13}%
\makeatletter
\providecommand \@ifxundefined [1]{%
 \@ifx{#1\undefined}
}%
\providecommand \@ifnum [1]{%
 \ifnum #1\expandafter \@firstoftwo
 \else \expandafter \@secondoftwo
 \fi
}%
\providecommand \@ifx [1]{%
 \ifx #1\expandafter \@firstoftwo
 \else \expandafter \@secondoftwo
 \fi
}%
\providecommand \natexlab [1]{#1}%
\providecommand \enquote  [1]{``#1''}%
\providecommand \bibnamefont  [1]{#1}%
\providecommand \bibfnamefont [1]{#1}%
\providecommand \citenamefont [1]{#1}%
\providecommand \href@noop [0]{\@secondoftwo}%
\providecommand \href [0]{\begingroup \@sanitize@url \@href}%
\providecommand \@href[1]{\@@startlink{#1}\@@href}%
\providecommand \@@href[1]{\endgroup#1\@@endlink}%
\providecommand \@sanitize@url [0]{\catcode `\\12\catcode `\$12\catcode
  `\&12\catcode `\#12\catcode `\^12\catcode `\_12\catcode `\%12\relax}%
\providecommand \@@startlink[1]{}%
\providecommand \@@endlink[0]{}%
\providecommand \url  [0]{\begingroup\@sanitize@url \@url }%
\providecommand \@url [1]{\endgroup\@href {#1}{\urlprefix }}%
\providecommand \urlprefix  [0]{URL }%
\providecommand \Eprint [0]{\href }%
\providecommand \doibase [0]{http://dx.doi.org/}%
\providecommand \selectlanguage [0]{\@gobble}%
\providecommand \bibinfo  [0]{\@secondoftwo}%
\providecommand \bibfield  [0]{\@secondoftwo}%
\providecommand \translation [1]{[#1]}%
\providecommand \BibitemOpen [0]{}%
\providecommand \bibitemStop [0]{}%
\providecommand \bibitemNoStop [0]{.\EOS\space}%
\providecommand \EOS [0]{\spacefactor3000\relax}%
\providecommand \BibitemShut  [1]{\csname bibitem#1\endcsname}%
\let\auto@bib@innerbib\@empty
\bibitem [{\citenamefont {Simon}\ \emph {et~al.}(1986)\citenamefont {Simon},
  \citenamefont {Seka}, \citenamefont {Goldman},\ and\ \citenamefont
  {Short}}]{Simon_1986}%
  \BibitemOpen
  \bibfield  {author} {\bibinfo {author} {\bibfnamefont {A.}~\bibnamefont
  {Simon}}, \bibinfo {author} {\bibfnamefont {W.}~\bibnamefont {Seka}},
  \bibinfo {author} {\bibfnamefont {L.~M.}\ \bibnamefont {Goldman}}, \ and\
  \bibinfo {author} {\bibfnamefont {R.~W.}\ \bibnamefont {Short}},\ }\href
  {\doibase 10.1063/1.865636} {\bibfield  {journal} {\bibinfo  {journal} {Phys.
  Fluids}\ }\textbf {\bibinfo {volume} {29}},\ \bibinfo {pages} {1704}
  (\bibinfo {year} {1986})}\BibitemShut {NoStop}%
\bibitem [{\citenamefont {Tanaka}\ \emph {et~al.}(1982)\citenamefont {Tanaka},
  \citenamefont {Goldman}, \citenamefont {Seka}, \citenamefont {Richardson},
  \citenamefont {Soures},\ and\ \citenamefont {Williams}}]{Tanaka_1982}%
  \BibitemOpen
  \bibfield  {author} {\bibinfo {author} {\bibfnamefont {K.}~\bibnamefont
  {Tanaka}}, \bibinfo {author} {\bibfnamefont {L.~M.}\ \bibnamefont {Goldman}},
  \bibinfo {author} {\bibfnamefont {W.}~\bibnamefont {Seka}}, \bibinfo {author}
  {\bibfnamefont {M.~C.}\ \bibnamefont {Richardson}}, \bibinfo {author}
  {\bibfnamefont {J.~M.}\ \bibnamefont {Soures}}, \ and\ \bibinfo {author}
  {\bibfnamefont {E.~A.}\ \bibnamefont {Williams}},\ }\href {\doibase
  10.1103/PhysRevLett.48.1179} {\bibfield  {journal} {\bibinfo  {journal}
  {Phys. Rev. Lett.}\ }\textbf {\bibinfo {volume} {48}},\ \bibinfo {pages}
  {1179} (\bibinfo {year} {1982})}\BibitemShut {NoStop}%
\bibitem [{\citenamefont {Shepard}\ \emph {et~al.}(1986)\citenamefont
  {Shepard}, \citenamefont {Tarvin}, \citenamefont {Berger}, \citenamefont
  {Busch}, \citenamefont {Johnson},\ and\ \citenamefont
  {Schroeder}}]{Shepard_1986}%
  \BibitemOpen
  \bibfield  {author} {\bibinfo {author} {\bibfnamefont {C.~L.}\ \bibnamefont
  {Shepard}}, \bibinfo {author} {\bibfnamefont {J.~A.}\ \bibnamefont {Tarvin}},
  \bibinfo {author} {\bibfnamefont {R.~L.}\ \bibnamefont {Berger}}, \bibinfo
  {author} {\bibfnamefont {G.~E.}\ \bibnamefont {Busch}}, \bibinfo {author}
  {\bibfnamefont {R.~R.}\ \bibnamefont {Johnson}}, \ and\ \bibinfo {author}
  {\bibfnamefont {R.~J.}\ \bibnamefont {Schroeder}},\ }\href {\doibase
  10.1063/1.865449} {\bibfield  {journal} {\bibinfo  {journal} {Phys. Fluids}\
  }\textbf {\bibinfo {volume} {29}},\ \bibinfo {pages} {583} (\bibinfo {year}
  {1986})}\BibitemShut {NoStop}%
\bibitem [{\citenamefont {Baldis}\ \emph {et~al.}(1989)\citenamefont {Baldis},
  \citenamefont {Young}, \citenamefont {Drake}, \citenamefont {Kruer},
  \citenamefont {Estabrook}, \citenamefont {Williams},\ and\ \citenamefont
  {Johnston}}]{Baldis_PRL_1989}%
  \BibitemOpen
  \bibfield  {author} {\bibinfo {author} {\bibfnamefont {H.~A.}\ \bibnamefont
  {Baldis}}, \bibinfo {author} {\bibfnamefont {P.~E.}\ \bibnamefont {Young}},
  \bibinfo {author} {\bibfnamefont {R.~P.}\ \bibnamefont {Drake}}, \bibinfo
  {author} {\bibfnamefont {W.~L.}\ \bibnamefont {Kruer}}, \bibinfo {author}
  {\bibfnamefont {K.}~\bibnamefont {Estabrook}}, \bibinfo {author}
  {\bibfnamefont {E.~A.}\ \bibnamefont {Williams}}, \ and\ \bibinfo {author}
  {\bibfnamefont {T.~W.}\ \bibnamefont {Johnston}},\ }\href {\doibase
  10.1103/PhysRevLett.62.2829} {\bibfield  {journal} {\bibinfo  {journal}
  {Phys. Rev. Lett.}\ }\textbf {\bibinfo {volume} {62}},\ \bibinfo {pages}
  {2829} (\bibinfo {year} {1989})}\BibitemShut {NoStop}%
\bibitem [{\citenamefont {Barr}, \citenamefont {Boyd},\ and\ \citenamefont
  {Mackwood}(1993)}]{Barr_1993}%
  \BibitemOpen
  \bibfield  {author} {\bibinfo {author} {\bibfnamefont {H.}~\bibnamefont
  {Barr}}, \bibinfo {author} {\bibfnamefont {T.}~\bibnamefont {Boyd}}, \ and\
  \bibinfo {author} {\bibfnamefont {A.}~\bibnamefont {Mackwood}},\ }\href
  {\doibase https://doi.org/10.1016/0375-9601(93)90295-B} {\bibfield  {journal}
  {\bibinfo  {journal} {Phys. Lett. A}\ }\textbf {\bibinfo {volume} {180}},\
  \bibinfo {pages} {435} (\bibinfo {year} {1993})}\BibitemShut {NoStop}%
\bibitem [{\citenamefont {Drake}\ and\ \citenamefont
  {Batha}(1991)}]{Drake_1991}%
  \BibitemOpen
  \bibfield  {author} {\bibinfo {author} {\bibfnamefont {R.~P.}\ \bibnamefont
  {Drake}}\ and\ \bibinfo {author} {\bibfnamefont {S.~H.}\ \bibnamefont
  {Batha}},\ }\href {https://doi.org/10.1063/1.859926} {\bibfield  {journal}
  {\bibinfo  {journal} {Phys. Fluids B}\ }\textbf {\bibinfo {volume} {3}},\
  \bibinfo {pages} {2936} (\bibinfo {year} {1991})}\BibitemShut {NoStop}%
\bibitem [{\citenamefont {Rose}\ and\ \citenamefont
  {Mounaix}(2011)}]{Rose_2011}%
  \BibitemOpen
  \bibfield  {author} {\bibinfo {author} {\bibfnamefont {H.~A.}\ \bibnamefont
  {Rose}}\ and\ \bibinfo {author} {\bibfnamefont {P.}~\bibnamefont {Mounaix}},\
  }\href {\doibase 10.1063/1.3581083} {\bibfield  {journal} {\bibinfo
  {journal} {Phys. Plasmas}\ }\textbf {\bibinfo {volume} {18}},\ \bibinfo
  {pages} {042109} (\bibinfo {year} {2011})}\BibitemShut {NoStop}%
\bibitem [{\citenamefont {Rosenberg}\ \emph {et~al.}(2020)\citenamefont
  {Rosenberg}, \citenamefont {Solodov}, \citenamefont {Seka}, \citenamefont
  {Follett}, \citenamefont {Myatt}, \citenamefont {Maximov}, \citenamefont
  {Ren}, \citenamefont {Cao}, \citenamefont {Michel}, \citenamefont
  {Hohenberger}, \citenamefont {Palastro}, \citenamefont {Goyon}, \citenamefont
  {Chapman}, \citenamefont {Ralph}, \citenamefont {Moody}, \citenamefont
  {Scott}, \citenamefont {Glize},\ and\ \citenamefont
  {Regan}}]{Rosenberg_2020}%
  \BibitemOpen
  \bibfield  {author} {\bibinfo {author} {\bibfnamefont {M.~J.}\ \bibnamefont
  {Rosenberg}}, \bibinfo {author} {\bibfnamefont {A.~A.}\ \bibnamefont
  {Solodov}}, \bibinfo {author} {\bibfnamefont {W.}~\bibnamefont {Seka}},
  \bibinfo {author} {\bibfnamefont {R.~K.}\ \bibnamefont {Follett}}, \bibinfo
  {author} {\bibfnamefont {J.~F.}\ \bibnamefont {Myatt}}, \bibinfo {author}
  {\bibfnamefont {A.~V.}\ \bibnamefont {Maximov}}, \bibinfo {author}
  {\bibfnamefont {C.}~\bibnamefont {Ren}}, \bibinfo {author} {\bibfnamefont
  {S.}~\bibnamefont {Cao}}, \bibinfo {author} {\bibfnamefont {P.}~\bibnamefont
  {Michel}}, \bibinfo {author} {\bibfnamefont {M.}~\bibnamefont {Hohenberger}},
  \bibinfo {author} {\bibfnamefont {J.~P.}\ \bibnamefont {Palastro}}, \bibinfo
  {author} {\bibfnamefont {C.}~\bibnamefont {Goyon}}, \bibinfo {author}
  {\bibfnamefont {T.}~\bibnamefont {Chapman}}, \bibinfo {author} {\bibfnamefont
  {J.~E.}\ \bibnamefont {Ralph}}, \bibinfo {author} {\bibfnamefont {J.~D.}\
  \bibnamefont {Moody}}, \bibinfo {author} {\bibfnamefont {R.~H.~H.}\
  \bibnamefont {Scott}}, \bibinfo {author} {\bibfnamefont {K.}~\bibnamefont
  {Glize}}, \ and\ \bibinfo {author} {\bibfnamefont {S.~P.}\ \bibnamefont
  {Regan}},\ }\href {\doibase 10.1063/1.5139226} {\bibfield  {journal}
  {\bibinfo  {journal} {Physics of Plasmas}\ }\textbf {\bibinfo {volume}
  {27}},\ \bibinfo {pages} {042705} (\bibinfo {year} {2020})}\BibitemShut
  {NoStop}%
\bibitem [{\citenamefont {Michel}\ \emph {et~al.}(2019)\citenamefont {Michel},
  \citenamefont {Rosenberg}, \citenamefont {Seka}, \citenamefont {Solodov},
  \citenamefont {Short}, \citenamefont {Chapman}, \citenamefont {Goyon},
  \citenamefont {Lemos}, \citenamefont {Hohenberger}, \citenamefont {Moody},
  \citenamefont {Regan},\ and\ \citenamefont {Myatt}}]{Michel_2019}%
  \BibitemOpen
  \bibfield  {author} {\bibinfo {author} {\bibfnamefont {P.}~\bibnamefont
  {Michel}}, \bibinfo {author} {\bibfnamefont {M.~J.}\ \bibnamefont
  {Rosenberg}}, \bibinfo {author} {\bibfnamefont {W.}~\bibnamefont {Seka}},
  \bibinfo {author} {\bibfnamefont {A.~A.}\ \bibnamefont {Solodov}}, \bibinfo
  {author} {\bibfnamefont {R.~W.}\ \bibnamefont {Short}}, \bibinfo {author}
  {\bibfnamefont {T.}~\bibnamefont {Chapman}}, \bibinfo {author} {\bibfnamefont
  {C.}~\bibnamefont {Goyon}}, \bibinfo {author} {\bibfnamefont
  {N.}~\bibnamefont {Lemos}}, \bibinfo {author} {\bibfnamefont
  {M.}~\bibnamefont {Hohenberger}}, \bibinfo {author} {\bibfnamefont {J.~D.}\
  \bibnamefont {Moody}}, \bibinfo {author} {\bibfnamefont {S.~P.}\ \bibnamefont
  {Regan}}, \ and\ \bibinfo {author} {\bibfnamefont {J.~F.}\ \bibnamefont
  {Myatt}},\ }\href {\doibase 10.1103/PhysRevE.99.033203} {\bibfield  {journal}
  {\bibinfo  {journal} {Phys. Rev. E}\ }\textbf {\bibinfo {volume} {99}},\
  \bibinfo {pages} {033203} (\bibinfo {year} {2019})}\BibitemShut {NoStop}%
\bibitem [{\citenamefont {Michel}(2023)}]{LPI_book}%
  \BibitemOpen
  \bibfield  {author} {\bibinfo {author} {\bibfnamefont {P.}~\bibnamefont
  {Michel}},\ }\href {https://books.google.co.il/books?id=anODzwEACAAJ} {\emph
  {\bibinfo {title} {Introduction to Laser-Plasma Interactions}}},\ Graduate
  Texts in Physics\ (\bibinfo  {publisher} {Springer International
  Publishing},\ \bibinfo {year} {2023})\BibitemShut {NoStop}%
\bibitem [{\citenamefont {Rosenbluth}(1972)}]{Rosenbluth_1972}%
  \BibitemOpen
  \bibfield  {author} {\bibinfo {author} {\bibfnamefont {M.~N.}\ \bibnamefont
  {Rosenbluth}},\ }\href {\doibase 10.1103/PhysRevLett.29.565} {\bibfield
  {journal} {\bibinfo  {journal} {Phys. Rev. Lett.}\ }\textbf {\bibinfo
  {volume} {29}},\ \bibinfo {pages} {565} (\bibinfo {year} {1972})}\BibitemShut
  {NoStop}%
\bibitem [{\citenamefont {Kruer}(2003)}]{Kruer_book}%
  \BibitemOpen
  \bibfield  {author} {\bibinfo {author} {\bibfnamefont {W.}~\bibnamefont
  {Kruer}},\ }\href {https://doi.org/10.1201/9781003003243} {\emph {\bibinfo
  {title} {The Physics Of Laser Plasma Interactions}}},\ Frontiers in Physics\
  (\bibinfo  {publisher} {CRC Press},\ \bibinfo {year} {2003})\BibitemShut
  {NoStop}%
\bibitem [{\citenamefont {Arber}\ \emph {et~al.}(2015)\citenamefont {Arber},
  \citenamefont {Bennett}, \citenamefont {Brady}, \citenamefont
  {Lawrence-Douglas}, \citenamefont {Ramsay}, \citenamefont {Sircombe},
  \citenamefont {Gillies}, \citenamefont {Evans}, \citenamefont {Schmitz},
  \citenamefont {Bell},\ and\ \citenamefont {Ridgers}}]{epoch}%
  \BibitemOpen
  \bibfield  {author} {\bibinfo {author} {\bibfnamefont {T.~D.}\ \bibnamefont
  {Arber}}, \bibinfo {author} {\bibfnamefont {K.}~\bibnamefont {Bennett}},
  \bibinfo {author} {\bibfnamefont {C.~S.}\ \bibnamefont {Brady}}, \bibinfo
  {author} {\bibfnamefont {A.}~\bibnamefont {Lawrence-Douglas}}, \bibinfo
  {author} {\bibfnamefont {M.~G.}\ \bibnamefont {Ramsay}}, \bibinfo {author}
  {\bibfnamefont {N.~J.}\ \bibnamefont {Sircombe}}, \bibinfo {author}
  {\bibfnamefont {P.}~\bibnamefont {Gillies}}, \bibinfo {author} {\bibfnamefont
  {R.~G.}\ \bibnamefont {Evans}}, \bibinfo {author} {\bibfnamefont
  {H.}~\bibnamefont {Schmitz}}, \bibinfo {author} {\bibfnamefont {A.~R.}\
  \bibnamefont {Bell}}, \ and\ \bibinfo {author} {\bibfnamefont {C.~P.}\
  \bibnamefont {Ridgers}},\ }\href {\doibase 10.1088/0741-3335/57/11/113001}
  {\bibfield  {journal} {\bibinfo  {journal} {Plasma Physics and Controlled
  Fusion}\ }\textbf {\bibinfo {volume} {57}},\ \bibinfo {pages} {113001}
  (\bibinfo {year} {2015})}\BibitemShut {NoStop}%
\end{thebibliography}%
